\documentclass[onecolumn]{aastex631}
\pdfoutput=1 %for arXiv submission
\usepackage{amsmath,amstext}
\usepackage[T1]{fontenc}
\usepackage{apjfonts} 
\usepackage[figure,figure*]{hypcap}
\usepackage{CJKutf8}
\usepackage{longtable}
\usepackage[section]{placeins}
\usepackage{CJKutf8}

\newcommand{\LIDO}{L$i$DO}
\newcommand{\fkoz}{$f_{\rm koz}$}

\shortauthors{Lawler et al.}

\begin{document}

\title{L$i$DO: Exploring the Stable Plutino Parameter Space}

\author[0000-0001-5368-386X]{Samantha~M. Lawler}
\affiliation{Campion College and the Department of Physics, University of Regina, Regina, SK S4S 0A2, Canada}

\author{Mark~Comte}
\affiliation{Campion College and the Department of Physics, University of Regina, Regina, SK S4S 0A2, Canada}

\author[0000-0003-4797-5262]{Rosemary~E. Pike}
\affiliation{Center for Astrophysics | Harvard \& Smithsonian; 60 Garden Street, Cambridge, MA, 02138, USA}

\author[0000-0003-4143-8589]{Mike~Alexandersen}
\affiliation{Center for Astrophysics | Harvard \& Smithsonian; 60 Garden Street, Cambridge, MA, 02138, USA}

\author[0000-0001-7244-6069]{Ying-Tung Chen (\begin{CJK*}{UTF8}{bkai}陳英同\end{CJK*})}
%(\Chi{"96}{"73}\Chi{"82}{"F1}\Chi{"54}{"0C})}
\affiliation{Institute of Astronomy and Astrophysics, Academia Sinica; 11F of AS/NTU Astronomy-Mathematics Building, No. 1 Roosevelt Rd., Sec. 4, Taipei 106319, Taiwan}

\author[0009-0004-7149-5212]{Cameron Collyer}
\affiliation{Center for Astrophysics | Harvard \& Smithsonian; 60 Garden Street, Cambridge, MA, 02138, USA}
\affiliation{Department of Astronomy, Boston University, 725 Commonwealth Avenue, Boston, MA 02215 USA}

\author{Matthew~Holman}
\affiliation{Center for Astrophysics | Harvard \& Smithsonian; 60 Garden Street, Cambridge, MA, 02138, USA}

\author[0000-0001-7032-5255]{J.~J. Kavelaars}
\affiliation{Herzberg Astronomy and Astrophysics Research Centre, National Research Council of Canada, 5071 West Saanich Rd, Victoria, BC V9E 2E7, Canada}
\affiliation{Department of Physics and Astronomy, University of Victoria, Victoria, BC, Canada}

\author{Lowell Peltier}
\affiliation{Department of Physics and Astronomy, University of Victoria, Victoria, BC, Canada}
\affiliation{Campion College and the Department of Physics, University of Regina, Regina, SK S4S 0A2, Canada}

\author{Cameron Semenchuck}
\affiliation{Campion College and the Department of Physics, University of Regina, Regina, SK S4S 0A2, Canada}
\affiliation{Department of Physics and Astronomy, University of Victoria, Victoria, BC, Canada}

\begin{abstract}

We present a publicly available, high-resolution, filled-parameter-space synthetic distribution of the Plutinos, trans-Neptunian Objects (TNOs) librating in the 3:2 mean-motion resonance with Neptune, with particular focus on the Plutinos simultaneously Kozai-librating. 
This synthetic distribution was built in preparation for results from the Large inclination Distant Objects (\LIDO) Survey, which pointed at locations on the sky where Kozai Plutinos are predicted to come to pericenter and are thus most easily detected in magnitude-limited surveys.
Although we do not expect the full stable parameter-space presented here to be populated with real TNOs, it provides a useful starting point for comparison with Neptune migration simulations and debiased observational results.  
Our new stable parameter space synthetic distribution of fictitious Plutinos is consistent with previous works, and we build on past results by focusing on the behavior of Kozai Plutinos over 4~Gyr integrations.
We find that 95\% of 4~Gyr stable Kozai Plutinos remain in the same $\omega$-libration island for the entire integration. 
This provides an interesting diagnostic opportunity: any asymmetry in the true number of 4~Gyr stable Kozai Plutinos in the two $\omega$-libration islands must be caused by the details of emplacement during giant planet migration.  
Through analysis of previously published Neptune migration models, we show that the intrinsic fraction of Plutinos captured into Kozai depends on Neptune's migration speed and mode.
Combining the filled-parameter-space synthetic distribution with future migration simulations and the results of the carefully characterized LiDO survey will enable interpretation of the intrinsic orbital distribution of the Kozai and non-Kozai Plutinos.

\end{abstract}

\section{Introduction}

Trans-Neptunian Objects (TNOs) that orbit within the 3:2 mean-motion resonance with Neptune are named Plutinos, after Pluto, the first known 3:2 resonator.  
Pluto was shown to be librating in this resonance in the very early days of numerical orbital integration \citep{Cohen1964}, and Pluto's eccentric, inclined, resonant orbit provided the first clues to the migration history of the giant planets in our Solar System \citep{Malhotra1993}.

The 3:2 resonance is likely the most populated resonance in the Kuiper Belt \citep{Gladman2012,Volk2016,Alexandersen2016}, and also one of the easiest resonances to study due to its relatively close location within the Kuiper Belt, with a semi-major axis of $a\sim39.4$~AU and pericenters close or even interior to Neptune's semimajor axis at $a\sim30.1$~AU.
Like all objects in the distant Solar System, Plutinos are observed in reflected light, thus observers are strongly biased toward discovering TNOs close to pericenter, where they are significantly brighter than during the rest of their orbit.
Plutinos and other resonant TNOs are bound by the resonant condition to come to pericenter at specific points on the sky relative to Neptune, causing very specific observation biases within this population \citep[e.g.][]{Lawler2013}.  
The position on the sky where Plutinos come to pericenter relative to Neptune (averaged over time) is centered on 90$^{\circ}$ ahead of and behind Neptune along the ecliptic plane.%, at the ``ortho-Neptune points.''  

The Plutinos are very well-studied theoretically, with many works mapping the extent and long-term stability of this population, even when only a handful of real Plutinos were known \citep{Morbidelli1997,Nesvorny2000}. These models were able to be more fully tested as more Plutinos were discovered \citep[e.g.][]{Tiscareno2009,Li2014}.
The Plutinos include a large sub-population that is simultaneously in the Kozai resonance \citep{Kozai1962,Lidov1962} \footnote{
This particular secular resonance is sometimes called the ``von Zeipel-Lidov-Kozai resonance'' \citep{Ito2019}, or more properly a ``periodic orbit of the third kind'' \citep{Malhotra1997} or ``$g$ libration'' \citep{Malhotra2023}, but we will refer to it throughout this work as ``Kozai'' for brevity.
}; Pluto itself is in this orbital resonance-in-a-resonance \citep{Williams1971}.
In an averaged system, in which the orbits of the planets (or other perturbers) are circularly symmetric, the Kozai resonance causes an exchange between inclination $i$ and eccentricity $e$ that conserves the $z$-component of angular momentum, and causes the argument of pericenter $\omega$ to librate around 90$^{\circ}$ or 270$^{\circ}$ with a period of a few Myr, rather than precessing as is typical for non-Kozai TNOs.
This causes some observational biases that are distinct from those for non-Kozai Plutinos: Kozai Plutinos come to pericenter and are thus most easily detected at relatively high ecliptic latitudes \citep{Lawler2013}.
TNO surveys have typically focused on discovery near the ecliptic plane, and thus the TNOs listed in e.g. the Minor Planet Center (MPC) database and discovered in surveys such as the Canada-France Ecliptic Plane Survey \citep[CFEPS;][]{Petit2011}, the Deep Ecliptic Survey \citep{Adams2014}, and the Outer Solar System Origins Survey \citep[OSSOS;][]{Bannister2018} are biased against discovering Kozai Plutinos.
Several surveys with very deep limits \citep[e.g.][]{Smotherman2024,Fraser2023} or very wide sky coverage \citep[e.g.][]{Schwamb2023} are in progress now or will be soon, and a new stability map will be helpful for understanding the properties of newly discovered Plutinos in these surveys.

In this work we present a new high-resolution, publicly-available, filled-parameter-space synthetic distribution of the stable Plutinos, paying careful attention to the orbital properties of Kozai Plutinos.  
As described in detail in Section~\ref{sec:sims}, our filled-parameter-space synthetic distribution is tens of thousands of simulated particles randomly distributed across the full range of possible $a$, $e$, $i$, and angular orbital elements that librate in the 3:2 mean-motion resonance with Neptune for 4~Gyr.
These simulations are not intended to reproduce the current Plutinos, but to explore the full stable parameter space where objects could remain in the 3:2 mean-motion resonance with Neptune for the age of the Solar System.
This synthetic distribution is useful for comparison with the output from Kuiper Belt emplacement simulations, to learn which parts of Plutino parameter space were populated during giant planet migration, and which remain empty despite long-term stability.
We build on previous stable Plutino models, and see the same broad structures as presented previously in the literature \citep[e.g.][]{Morbidelli1997,Nesvorny2001,Tiscareno2009}. %,Li2014}.
In Section~\ref{sec:resonant} we discuss the basic dynamics of the 3:2 resonance and the Kozai resonance within it, as well as our dynamical integrations and classification.
In Section~\ref{sec:classified} we describe the properties of the filled-parameter-space Plutino synthetic distribution created by our simulations, including $a-e-i$ distributions and libration amplitudes within the 3:2 and Kozai resonances.
In particular, Section~\ref{sec:kozai} gives a detailed description of the behaviour of the Kozai Plutinos, both those that are 4~Gyr stable and those that are in the Kozai resonance for only part of the 4~Gyr integration.
In Section~\ref{sec:Discussion}, we discuss implications of these results for published and future giant planet migration models as well as future TNO discovery surveys.

%%%%%%%%%%%%%%%%%%%%%%%%%%%%%%%%%%%%%%%%%%%%%%%%%%%%%%%%%5
\section{Resonant Dynamics and Dynamical Classification}
\label{sec:resonant}

To diagnose if an object is in a mean-motion resonance with Neptune, the resonant angle, $\phi$, must be examined over the course of a long (at least several Myr) $n$-body integration. The resonant angle for Plutinos in the 3:2 mean-motion resonance with Neptune is
\begin{equation}
    \phi_{3:2}=3\lambda - 2\lambda_{\rm N} - \boldsymbol{\varpi} \label{eq:phi}
\end{equation} 
\\
Here, the longitude of pericentre $\boldsymbol{\varpi}$ is defined as the sum of the argument of pericenter $\omega$ and the longitude of the ascending node $\Omega$.  
$\lambda$ and $\lambda_{\rm N}$ are the mean longitudes for the object and for Neptune, respectively. 
Mean longitude $\lambda$ is defined as the sum of $\boldsymbol{\varpi}$ and the mean anomaly $\mathcal{M}$, which is the time since the last pericenter multiplied by the mean motion, $2\pi/ P$, where $P$ is orbital period. 
Objects in the 3:2 resonance (Plutinos) will have their $\phi_{3:2}$ librate about 180$^{\circ}$ instead of circulating.

The Kozai resonance within the Plutinos affects the behavior of the argument of pericenter $\omega$ over time, and occurs only at moderately high inclinations and eccentricities \citep[e.g.][]{Morbidelli1997,Wan2007}. 
Analogous to $\phi_{3:2}$ resonant behavior, the Kozai resonance forces $\omega$ to librate around a value over the course of the integration instead of circulating, most stably librating around 90$^{\circ}$ or 270$^{\circ}$.
Unstable, temporary Kozai libration can also occur around $\omega=0^{\circ}$ or 180$^{\circ}$ (see Section~\ref{sec:zero180}).
Due to conservation of the $z$-component of angular momentum, the Kozai resonance also forces $e$ and $i$ to be anti-correlated with the same frequency as the $\omega$ libration (few Myr).
However, other effects can cause this anti-correlated $e$-$i$ behaviour, so that alone does not prove Kozai resonance: $\omega$-libration is the defining feature to search for in orbital integrations to diagnose Kozai behaviour.

\subsection{The \LIDO\ Survey: Motivation for this Theoretical Stability Study}

The Large Inclination Distant Objects (\LIDO) Survey is a TNO discovery and tracking project carried out on the Canada-France-Hawaii Telescope (CFHT) using the wide-field MegaCam imager, with observations carried out between the 2020A and 2023A semesters\footnote{Joint program between Canada and Taiwan, %co-PIs Samantha Lawler and Charles Ying-Tung Chen, 
programs 20AC02, 20AT02, 20BC19, 20BT05, 21AC13, 21AT04, 21BC03, 21BT03, 22AC21, 22AT11, 22BC18 and 23AC16 %\citep{Alexandersen2023}
}.
This survey was designed to be well-characterized, with the goal of tracking all discovered TNOs and measuring detection efficiencies and depths in all fields, in order to facilitate bias-corrected comparison between models and observations \citep[e.g.][]{Lawler2018} and to be used together with the OSSOS++ federation of surveys \citep{Petit2011,Alexandersen2016,Petit2017,Bannister2018}.

\begin{figure}%[!hp]
    \centering \includegraphics[width=0.8\textwidth]{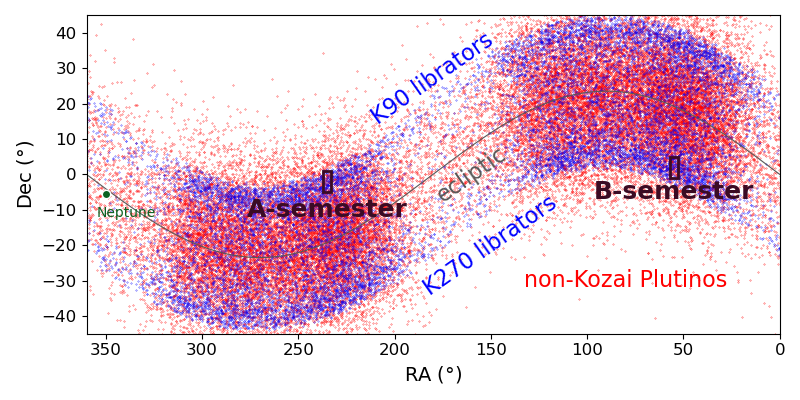}
    \caption{A snapshot of the predicted positions of a flux-limited set of simulated Plutinos on the sky, generated using a toy model version of the expected \LIDO\ survey sensitivity and an orbital element distribution model based on debiased measurements from \citet{Lawler2013} and \citet{Alexandersen2016}. Kozai Plutinos are shown with blue points and non-Kozai Plutinos in red.  The ecliptic and position of Neptune at the midpoint of the survey are also marked.  Our CFHT program targeted the highest predicted on-sky density of K90 librators in the A-semester block and K270 librators in the B-semester block.}  
    \label{fig:pointings}
\end{figure}

One of the main scientific motivations behind \LIDO\ was to discover more Kozai Plutinos. 
Because the Kozai resonance confines $\omega$ to librate around 90$^{\circ}$ (which we refer to as ``K90 librators'') or 270$^{\circ}$ (``K270 librators''), Kozai Plutinos always reach pericenter off the ecliptic plane.
This means that Kozai Plutinos will on average be brightest and easiest to detect at least 10-30 degrees away from the ecliptic plane \citep{Lawler2013}, and since most TNO surveys have focused on discoveries near the ecliptic plane where the overall population of TNOs has the highest on-sky densities, these ecliptic surveys are strongly biased against discovering Kozai Plutinos.
The two $\omega$-libration centers of the Kozai resonance mean there are separate areas of best detectability on the sky: objects in each island will come to pericenter exclusively above (K90 librators) or exclusively below (K270 librators) the ecliptic plane due to libration of the argument of pericenter $\omega$ \citep{Lawler2013}.
As shown in Figure~\ref{fig:pointings}, the \LIDO\ survey was designed to optimize Kozai Plutino detections, by targeting two off-ecliptic blocks, one on the area where the K90 librators are most easily detectable, and one where the K270 librators are most detectable.
These locations were determined based on the results of previous modeling efforts \citep{Lawler2013,Alexandersen2016}. However, the inclination width of Kozai Plutinos is not well constrained due to the small numbers of Kozai Plutinos detected in previous surveys, and measuring this inclination distribution robustly is a goal of the \LIDO\ survey.
As a result, the \LIDO\ survey discovery blocks were designed to optimize detection at a range of ecliptic latitudes.
Combining the Kozai and non-Kozai Plutino detections from \LIDO\ with detections from previous well-characterized surveys will provide a sufficiently large sample of Kozai and non-Kozai Plutinos to constrain the intrinsic fraction of Plutinos in each of the Plutino sub-populations.

The $\LIDO$ survey has completed observing and analyses of the orbital properties of the discovered sample are forthcoming.
Data were acquired from CFHT MegaCam in 2020A through 2023A, and additional astrometric data was acquired from Magellan Baade, Gemini North, and archival images for objects which required additional sampling.
$\LIDO$ discovered 141 objects, 125 of which are characterized, meaning that their discovery likelihood can be quantified with sufficient accuracy for statistical studies.
Some preliminary results from the survey are shown here, however, additional work is ongoing to finalize the orbits and classifications of the discoveries and to interpret their significance.

%%%%%%%%%%%%%%%%%%%%%%%%%%%%%%%%%%%%%%%%%%%%%%%%%%%%%%%%%
\subsection{Simulation Design}
\label{sec:sims}

Our goal in this work is to develop a filled-parameter-space synthetic distribution of the Plutinos for exploring survey observation biases and as a starting point for testing giant planet migration models.
We build on previous efforts \citep[e.g.,][]{Nesvorny2000, Tiscareno2009,Li2014}, and in order to facilitate future research, our filled-parameter-space synthetic distribution is publicly available\footnote{The end-state osculating orbital elements along with libration centers and libration amplitudes for Kozai and non-Kozai Plutinos are available at \url{https://www.canfar.net/citation/landing?doi=23.0028}}. 
These stability models will be used as starting point for understanding observational biases in \LIDO\ (Alexandersen et al. in prep.), which focused on discovery of high inclination TNOs.  Thus, in our simulation design, we emphasize including a significant number of initially high-$i$ particles. The higher resolution is important as real TNOs have associated orbital uncertainties which may result in them spanning multiple classifications within their possible orbital elements when these uncertainties are included. A model which indicates which behaviors we should expect to find for each object before individual object integrations are run provides a useful test for whether our real object integrations are sufficiently well-sampled.  This synthetic distribution can also be used to determine which parts of Plutino stable parameter space are filled by Neptune migration simulations \citep[e.g.,][]{Hahn2005,Levison2008,Brasser2013,Kaib2016,Nesvorny2016,Balaji2023}, and which parts remain devoid of known Plutinos despite long-term stability.

In order to identify the stable and unstable Plutino parameter space, we performed a large set of orbital integrations on the cluster facilities of the Canadian Advanced Network for Astronomy Research (CANFAR).
We used the WHFAST module \citep{Rein2015} within REBOUND \citep{rein2012} to integrate the four giant planets and an initial one million test particles with a 0.5~year integration timestep.
The four giant planets were initialized at their orbital positions on 1 January 2021, and the mass of the four terrestrial planets was added to the mass of the Sun.
Test particles had their initial orbital elements chosen randomly within the following range of orbital elements, which we verified does indeed cover all of stable Plutino parameter space (adding particles outside this parameter space resulted in no additional stable librators): $39.0$~AU~$<a<39.8$~AU, $0<e<0.5$, $0^{\circ}<i<90^{\circ}$.
Because no retrograde Plutinos are known, we only included inclinations up to 90$^{\circ}$.
The angular orbital elements: longitude of the ascending node $\Omega$, argument of pericenter $\omega$, and mean longitude $\mathcal{M}$, were chosen for each test particle randomly between 0$^{\circ}$-360$^{\circ}$.
We generated one million test particles, and initial short integrations were performed to search for libration of the resonant angle $\phi_{3:2}$.  We retained only the librating test particles for longer integrations, resulting in 217,560 test particles kept after the first 10~Myr simulation for further integrations.
The integrations were performed incrementally with an initial integration time of 10~Myr where the non-librating particles were removed and the stably librating particles were kept to be integrated for a longer time period. 
This process was repeated with integration times continuing to 100, 500 and 900~Myr until roughly a hundred thousand particles remained stable.
Those remaining particles had their integrations continued to 4 billion years to determine the long term stability of the Plutino parameter space.
For the final 4~Gyr integration a writing cadence of ten million years was used to ensure sample resolution to confirm stability.

\begin{figure}%[!hp]
    \centering
    \includegraphics[width=0.8\textwidth]{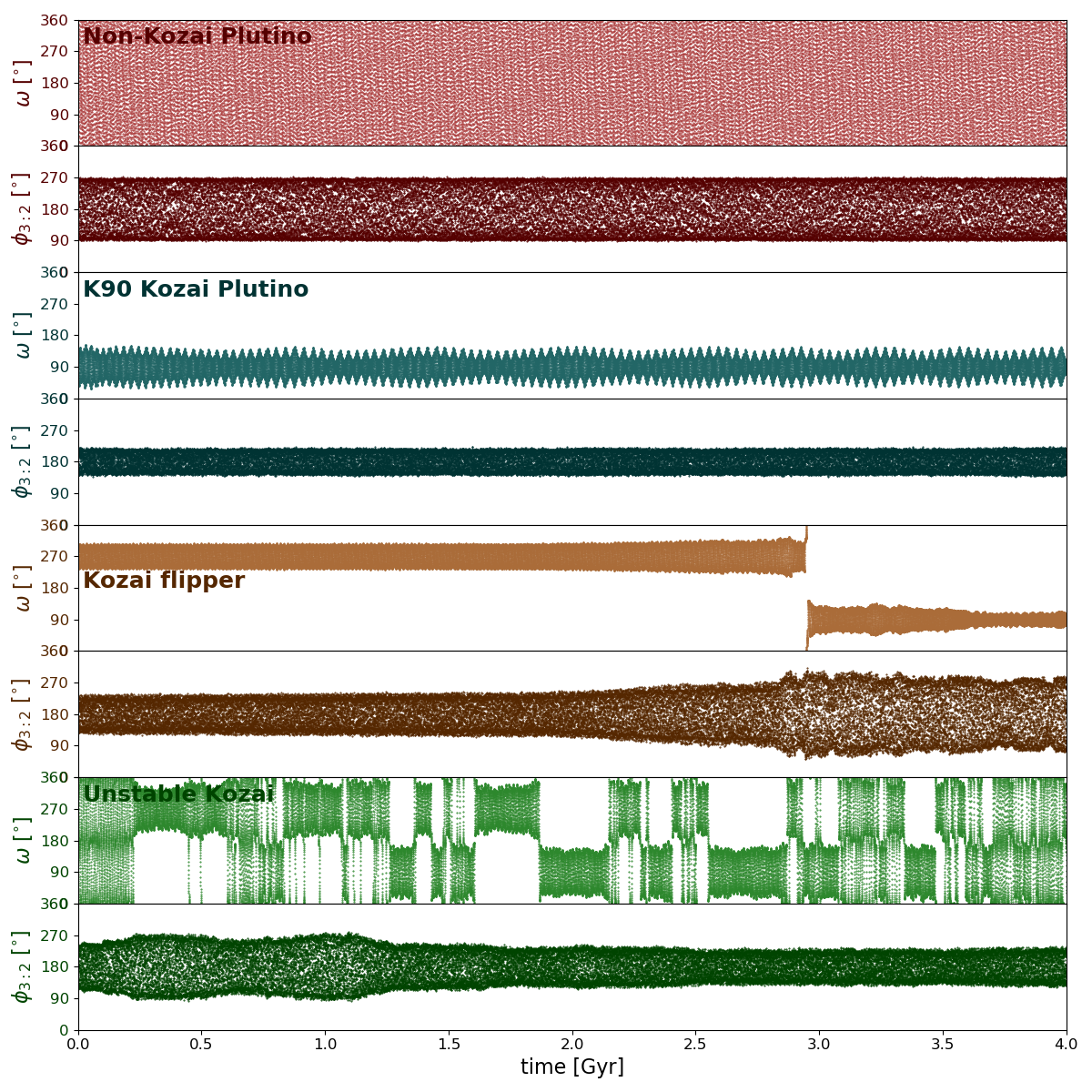}
    \caption{The argument of pericenter $\omega$ and the resonant angle $\phi_{3:2}$ plotted over 4~Gyr simulations (sampled at 10~Myr intervals) for four different test particles showing the range of behaviors identified within the 3:2 resonance.  Each test particle is labeled with the classification determined by our classification script.   
    }  
    \label{fig:plutino_integrations}
\end{figure}

The classification of the test particles into Plutinos and Kozai Plutinos requires a careful analysis of the integration output.
The algorithm to determine whether the particles are in a 3:2 mean-motion resonance with Neptune first calculates the 3:2 resonant angle (Equation~\ref{eq:phi}) at all time steps, then checks whether the resonant angle is confined (librating) or takes on all values 0$^{\circ}$-360$^{\circ}$ (circulating). 
Because libration amplitudes and modes can change over the course of a long integration, the algorithm examines the resonant angle behavior in time windows that are each 1/10 the integration length. 
Within each time window, the resonant angles are placed into 15$^{\circ}$ bins. 
We settled on 15$^{\circ}$ bins as a compromise between detecting the largest libration amplitude Plutinos and not adding many false positives where time-sample beating or random gaps cause no points to land in a given bin.  
Few very large libration amplitude Plutinos have been discovered, though surveys like CFEPS and OSSOS should have been able to detect them \citep[e.g.][]{Volk2016}, so we are confident that 15$^{\circ}$ bins will not exclude a part of parameter space that is likely to be populated.
If any $\phi_{3:2}$ bins remain empty, the particle is librating in the 3:2 resonance during that time-window.
If the particle is librating in all ten of the time windows, then the test particle is classified as a Plutino. 
Several example 4~Gyr integrations of stable Plutino are shown in Figure~\ref{fig:plutino_integrations}.
All Plutino test particles are then checked for Kozai resonance.
The same algorithm is followed, but now checking for empty 15$^{\circ}$ bins in argument of pericenter $\omega$ over each time-window.
If the particle's $\omega$ is librating in all of the ten windows, then the particle is classified as a Kozai Plutino. 
Example $\phi_{3:2}$ and $\omega$ evolutions for a 4~Gyr integration of a stable Kozai Plutino and non-Kozai Plutino are shown in Figure~\ref{fig:plutino_integrations}.
The orbital distribution of these classified resonators provides a map of the stable parameter space in the 3:2 resonance.

%%%%%%%%%%%%%%%%%%%%%%%%%%%%%%%%%%%%%%%%%%%%%%%%%%%%%%%%%
\section{Stable Plutino Parameter Space}
\label{sec:classified}

\begin{figure}%[!hp]
    \includegraphics[width=1.0\textwidth]{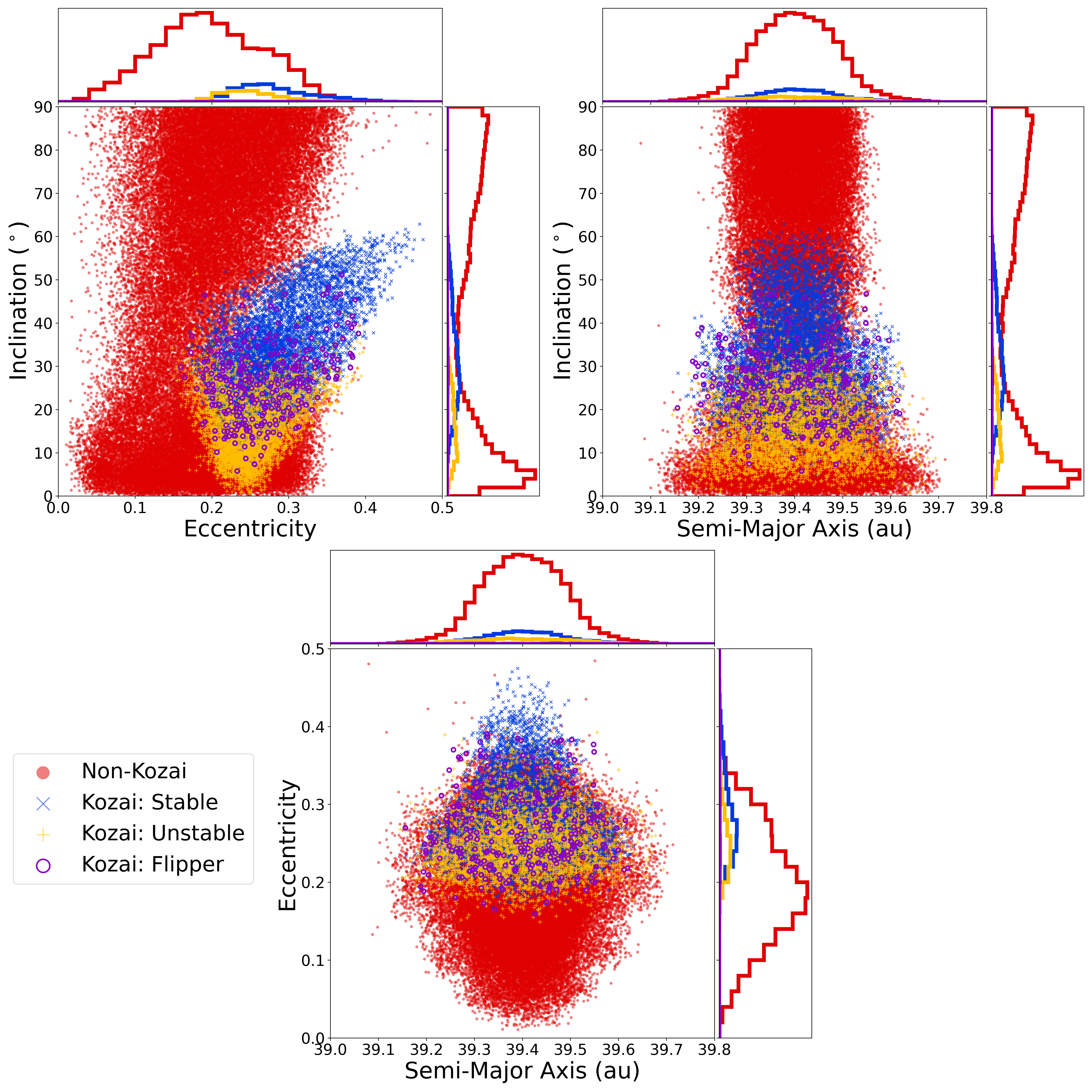}
    \caption{Osculating end-state orbital elements for 4~Gyr stable Plutinos, with non-Kozai Plutinos in red, stable Kozai Plutinos in blue, unstable Kozai oscillators in yellow, and Kozai flippers in purple. 
    Histograms along the x and y-axes of each plot show the relative number densities of each set of objects across each parameter space cut. Note that at the central $a$ value for the 3:2 resonance, $e>0.24$ will result in a pericenter sunward of Neptune's orbit.
    }  
    \label{fig:kozai_flipaei}
\end{figure}

\begin{figure}%[!hp]
    \includegraphics[width=0.5\textwidth]{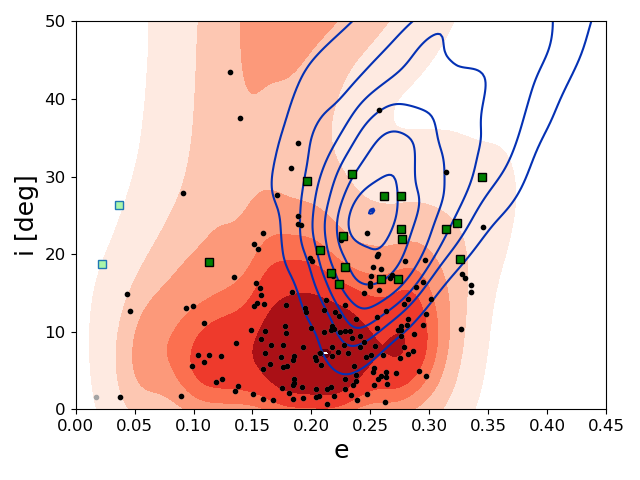}
    \includegraphics[width=0.5\textwidth]{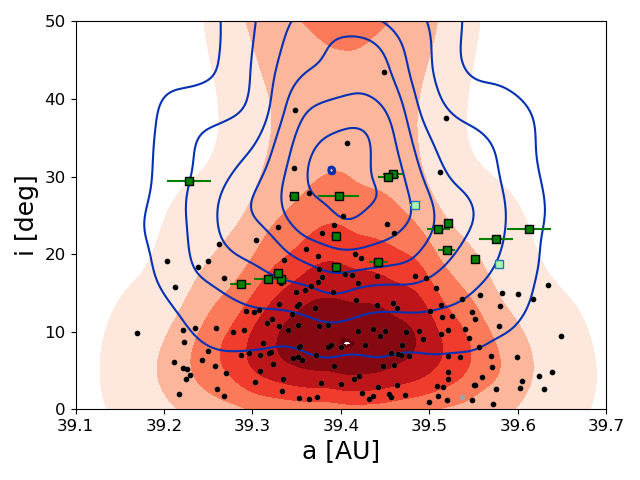} 
    \flushright \includegraphics[width=0.5\textwidth]{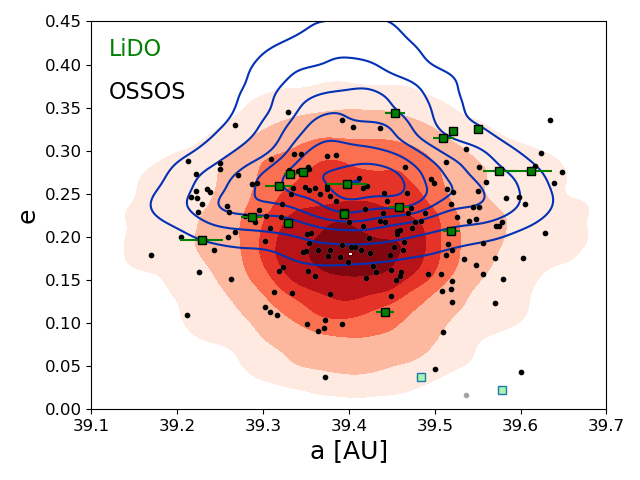}
    \caption{Nested contour lines or shades show the relative density of stable Plutinos and Kozai Plutinos from Figure~\ref{fig:kozai_flipaei} in $a$, $e$, and $i$: red contour shades show non-Kozai Plutinos and blue contours show the Kozai Plutinos.  Contours enclose 0.1\%, 10\%, 30\%, 50\%, 70\%, 90\%, and 99\% of synthetic distribution points. Overlaid are real TNOs from characterized surveys that are close to $a=39.4$~AU. Those detected by the \LIDO\ Survey are dark green squares (Plutinos) or light green squares (non-Plutinos), those previously detected by OSSOS++ are black (Plutinos) or grey points (non-Plutinos), the orbital uncertainties on OSSOS++ TNOs are smaller than the point size.  
    The \LIDO\ classifications are preliminary, but have sufficiently long arcs and use the same methodology as this work.
    The $i$-range of these plots is trimmed as while the high-$i$ parameter space is stable, no real TNOs have been discovered in this part of the resonance.
    } 
    \label{fig:realplut_aei}
\end{figure}

Figure~\ref{fig:kozai_flipaei} shows the osculating end-state orbital elements for 69,626 test particles that were librating in the 3:2 resonance for 4~Gyr of integration with the four giant planets, covering the entire (prograde) stable Plutino parameter space.
Kozai and non-Kozai Plutinos are denoted by different colours, with 7,529 test particles (10.8\% of all Plutino test particles) librating stably in Kozai for the entire 4~Gyr. 
These parameter space plots reproduce many of the features, such as both low and high-$i$ clumps of non-Kozai Plutinos, the $e$-$i$ dependence of the Kozai Plutinos, and stable region concentrated near $a=39.4$~AU, as previously discussed in the literature \citep[e.g.,][]{Nesvorny2000,Tiscareno2009}, but now at higher resolution.  
Figure~\ref{fig:realplut_aei} shows the stable model Kozai and non-Kozai Plutinos as density contours, for easier comparison to known Plutinos.

Figures~\ref{fig:kozai_flipaei} and \ref{fig:realplut_aei} map out the stably librating Plutino parameter space. The widest part of the stable non-Kozai Plutino space in $a$ occurs at an eccentricity of approximately 0.2, and fairly low inclinations of $\sim$5-8$^{\circ}$, while the widest part of Kozai Plutino stable space in $a$ occurs at slightly higher $e\sim0.25$ and much higher $i\sim30^{\circ}$.  The highest density portion of non-Kozai Plutino stable space is at $e\sim0.2$ and $i\sim5^{\circ}$, while the highest densities of Kozai Plutino stable parameter space sit at higher $e\sim0.25$ and $i\sim25^{\circ}$.

The comparison between the long-term stable parameter space and the known Plutinos can provide insight into the expected orbital behaviour of these objects and identify which regions of the stable parameter space were populated during the giant planet migration era.
In addition to the stable Plutino parameter space, Figure~\ref{fig:realplut_aei} also shows real TNOs from OSSOS++ (black and grey points) and \LIDO\ (green squares).  
The black points (nearly all of the OSSOS++ points) have been dynamically classified as Plutinos \citep{Petit2011,Alexandersen2016,Bannister2018}.
\LIDO-discovered TNOs near $a=39.4$~AU (green squares) have preliminary dynamical classifications nearly identical to the classification methodology used for the test particles in this work. Those which are librating in the 3:2 resonance are shown in dark green, non-Plutinos are in light green. Final astrometric measurements were obtained in late 2023 and will be discussed in an upcoming paper.
Note that because \LIDO\ was entirely off-ecliptic, only TNOs with inclinations above $i\sim15^{\circ}$ were discoverable (very obvious in the inclination panels of Figure~\ref{fig:realplut_aei}).

As discussed in \citet{Lan2019}, there is clearly a high-inclination stable component to the Plutino parameter space ($i\gtrsim$45$^{\circ}$).  But just because this parameter space is stable does not mean that it is populated; the highest inclination known classified Plutino has $i=43^{\circ}$ \citep{Bannister2018}.  
By focusing on discoveries well off the ecliptic, \LIDO\ had more potential to discover high-inclination Plutinos than OSSOS++, which was primarily on-ecliptic.  
However, as seen in Figure~\ref{fig:realplut_aei}, the highest inclination potential Plutinos detected in \LIDO\ were near $i\sim30^{\circ}$, even though the survey was sensitive to Plutinos at even higher inclinations.
We can tentatively say that the lack of very high inclination observed Plutinos in the preliminary \LIDO\ survey, despite abundant 4~Gyr stable parameter space, shows that this high inclination parameter space was likely never populated, including during Neptune migration and subsequent resonant sticking \citep{Lykawka2007b}.
As discussed in the literature, this may rule out some migration scenarios where particles are captured in large numbers into stable high-inclination Plutino orbits \citep[e.g.,][]{Volk2019}.
However, cautious interpretation is merited, as we are comparing simulations with observationally biased TNO detections, and the lack of discoveries is not always indicative of a lack of intrinsic objects.  
We note that previously published well-characterized surveys were sensitive to high inclination Plutinos and would have detected them if the population was significant \citep[in particular][]{Alexandersen2016,Petit2017}.
\citet{Volk2016} provides an intial OSSOS++ debiased Plutino model, and upcoming \LIDO\ analyses will use the Survey Simulator methodology \citep{Lawler2018} to provide upper limits on the inclination distribution of real Plutinos that are possible within our survey's observational limits.

Figures~\ref{fig:kozai_flipaei} and \ref{fig:realplut_aei} make it clear that out of the possible Plutino stable parameter space, Kozai Plutinos dominate at the highest stable eccentricities, moderate inclinations, and effectively the entire range of semimajor axes where non-Kozai Plutinos are stable. 
While a parameter space plot like this shows the relative populations of Kozai and non-Kozai Plutinos and could in theory be used to diagnose likely Kozai resonance for newly discovered TNOs, 
definitive Kozai diagnosis requires careful astrometric measurements over at least three oppositions and integration of clones within the astrometric errors to determine behaviour over $\sim$30~Myr.
This analysis is forthcoming for \LIDO\ TNO detections.

In the sections below we discuss different aspects of the stable parameter space distribution, and why each may or may not be visible in the real, observable population of Plutinos.

%%%%%%%%%%%%%%%%%%%%%%%%%%%%%%%%%%%%
\subsection{The Kozai fraction \fkoz}

One of the properties of Plutinos commonly measured in well-characterized surveys and in emplacement simulations is the fraction of Plutinos currently in Kozai, \fkoz.
Due to the on-sky and orbital observation biases, the values measured for \fkoz\ will differ widely depending on the particulars of any given discovery survey \citep[see ][for an extensive discussion of these effects]{Lawler2013}. 
\citet{Lawler2013} contains a summary of the studies up to that point, and more recently \citet{Balaji2023} uses a survey simulator to compare observational biases with emplacement models specifically for the Plutinos.
We do not expect the filled-parameter-space synthetic distribution presented here to match the current state of the Plutinos, but this parameter space synthetic distribution is useful for applying cuts in orbital elements to show very simply how observational biases (e.g., surveys limited to the ecliptic plane and thus more likely to detect low-$i$ Plutinos, and the biases toward detecting higher eccentricity TNOs close to pericenter) will affect the measured value of \fkoz.  

Some simple cuts and the resulting values of \fkoz\ for the dynamically classified particles in the filled-parameter-space synthetic distribution are presented in Table~\ref{tab:fkoz}, highlighting typical simple biases in TNO discovery surveys.
As cuts are made at lower inclination upper limits, \fkoz\ decreases.  
Most TNO surveys will have this bias, as they typically focus on the ecliptic.
As cuts are made to higher eccentricity lower limits, \fkoz\ increases.  
Magnitude-limited TNO surveys are typically biased toward detection of higher eccentricity TNOs \citep[see, e.g.][]{Jones2006}, so this bias will also be present.
On-sky locations will also have a strong effect on the measured \fkoz\ value \citep{Lawler2013}, and all these biases must be disentangled simultaneously to measure the true value of \fkoz.
Different or more complicated sub-samples of the synthetic distribution, which can be made by downloading the orbital elements and classifications of the particles from this work, will provide a useful comparison sample for simulation work and real detections.

\begin{deluxetable}{c|ccc}
\tablecaption{\fkoz\ measured for different subsets of the Plutino synthetic distribution \label{tab:fkoz}}
\tablewidth{0pt}
\tablehead{\colhead{Subset} & Kozai Plutinos & non-Kozai Plutinos & \fkoz
}
\startdata
Total & 7529 & 62097 & 0.108 \\ \hline
$i<45^{\circ}$ & 6464 & 32913 & 0.164 \\
$i<30^{\circ}$ & 3704 & 29299 & 0.112 \\
$i<15^{\circ}$ & 329 & 20725 & 0.016 \\ \hline
$e>0.1$ & 7529 & 57131 & 0.116 \\
$e>0.2$ & 7409 & 30560 & 0.195 \\
$e>0.3$ & 2313 & 4825 & 0.324 
\enddata
%\tablecomments{}
\end{deluxetable}

%%%%%%%%%%%%%%%%%%%%%%%%%%%%%%%%%
\subsection{$\phi_{3:2}$-Libration and $\omega$-Libration Amplitude Distributions}

\begin{figure}%[!hp]
    \includegraphics[width=0.5\textwidth]{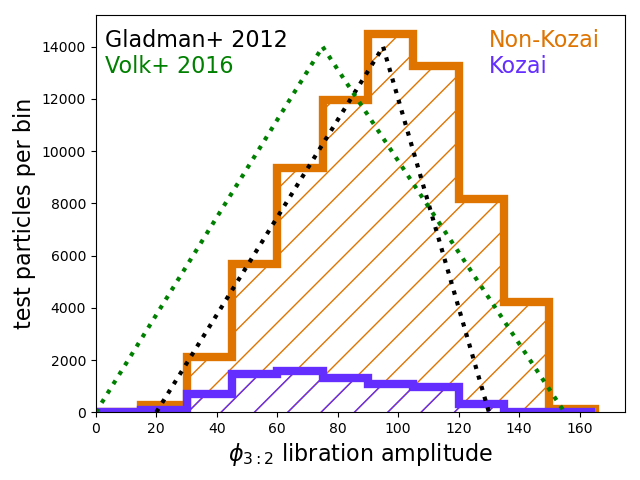}
    \includegraphics[width=0.5\textwidth]{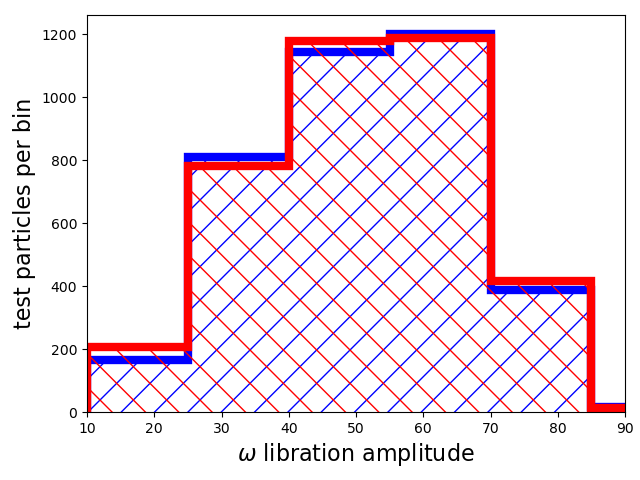}
    \caption{On the left is the distribution of $\phi_{3:2}$ for all 4~Gyr stable plutinos (Kozai and non-Kozai summed in a stacked histogram) in orange, with the Kozai Plutinos only in purple.  The difference in peak libration amplitude is likely a result of the $a$ and $e$ dependence of $\phi_{3:2}$ (see Figure~\ref{fig:libamps}).  The black and green dotted lines show the parametric distributions used in models based on debiased CFEPS \citep{Gladman2012} and OSSOS \citep{Volk2016} data, respectively, both of which used the same libration amplitude distributions for Kozai and non-Kozai Plutinos.  Right panel shows $\omega$-libration amplitudes for Kozai librators only - note that these amplitudes vary quite a bit over the course of the simulation, so large bins are used to show the overall distribution. K90 librators are in blue, K270 librators in red - there is no measurable difference between the distributions.
    }  
    \label{fig:libamphist}
\end{figure}

The left panel of Figure~\ref{fig:libamphist} shows the distribution of $\phi_{3:2}$-libration amplitudes for all 4~Gyr stable test particles.  As expected, this overall distribution matches decently with that presented in \citet{Nesvorny2000}, and somewhat (though not perfectly) matches the parametric models that have been used for Plutino orbital modeling and model-observation comparison in previous well-characterized surveys \citep[e.g.][]{Gladman2012,Alexandersen2016,Volk2016}. 
These libration amplitude distributions were used for both the Kozai and non-Kozai Plutinos, though the overall distributions are slightly different in this filled-parameter-space synthetic distribution.
It is not surprising that the match is not perfect, since the previously published $\phi_{3:2}$ distributions are debiased measurements, which will not necessarily be the same as the filled-parameter-space synthetic distribution.
For the Kozai Plutinos, we also roughly measure the $\omega$-libration amplitude distribution in a fairly coarse histogram (Figure~\ref{fig:libamphist}, right panel), due to the large variation in $\omega$ amplitude over the course of each simulation.
Our simulations show show very low ($\lesssim 20^{\circ}$) and very high ($>70^{\circ}$) libration amplitudes are not long-term stable. Low libration amplitudes should be stable theoretically, but in our simulations including all four giant planets, perturbations from the other planets quickly lead to higher libration amplitudes.

\begin{figure}%[!hp]
    \includegraphics[width=0.5\textwidth]{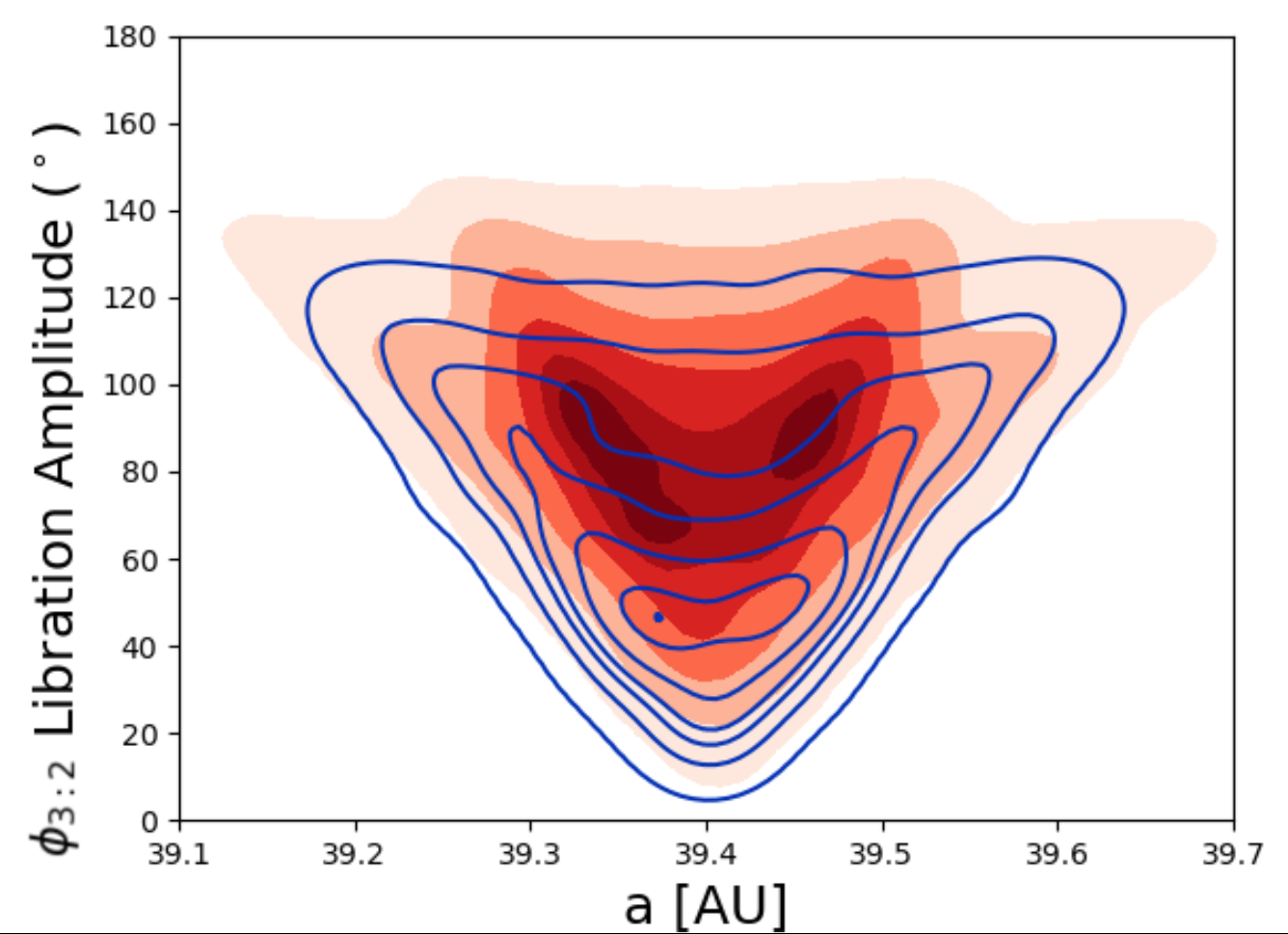}
    \includegraphics[width=0.5\textwidth]{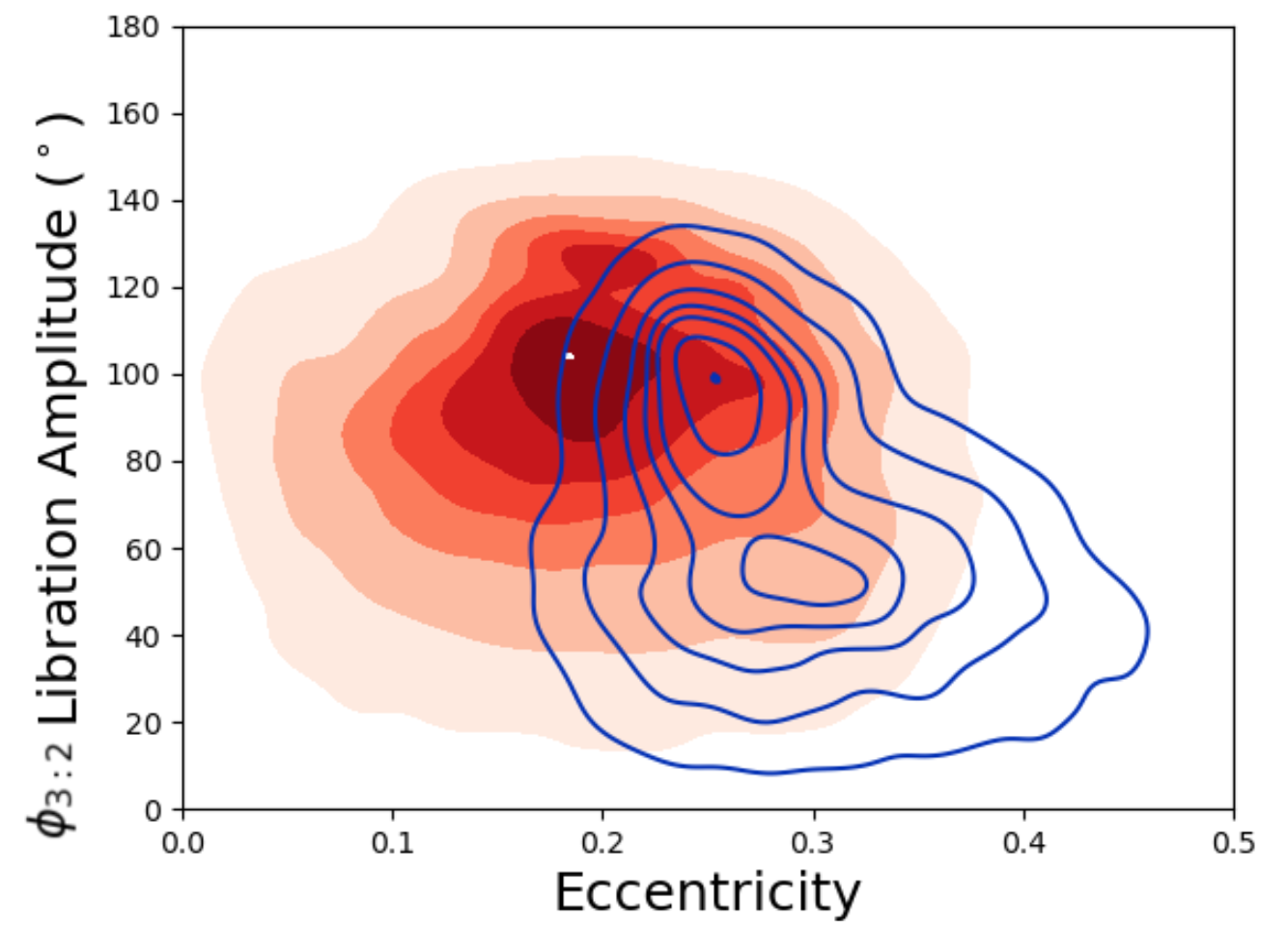}
    \includegraphics[width=0.5\textwidth]{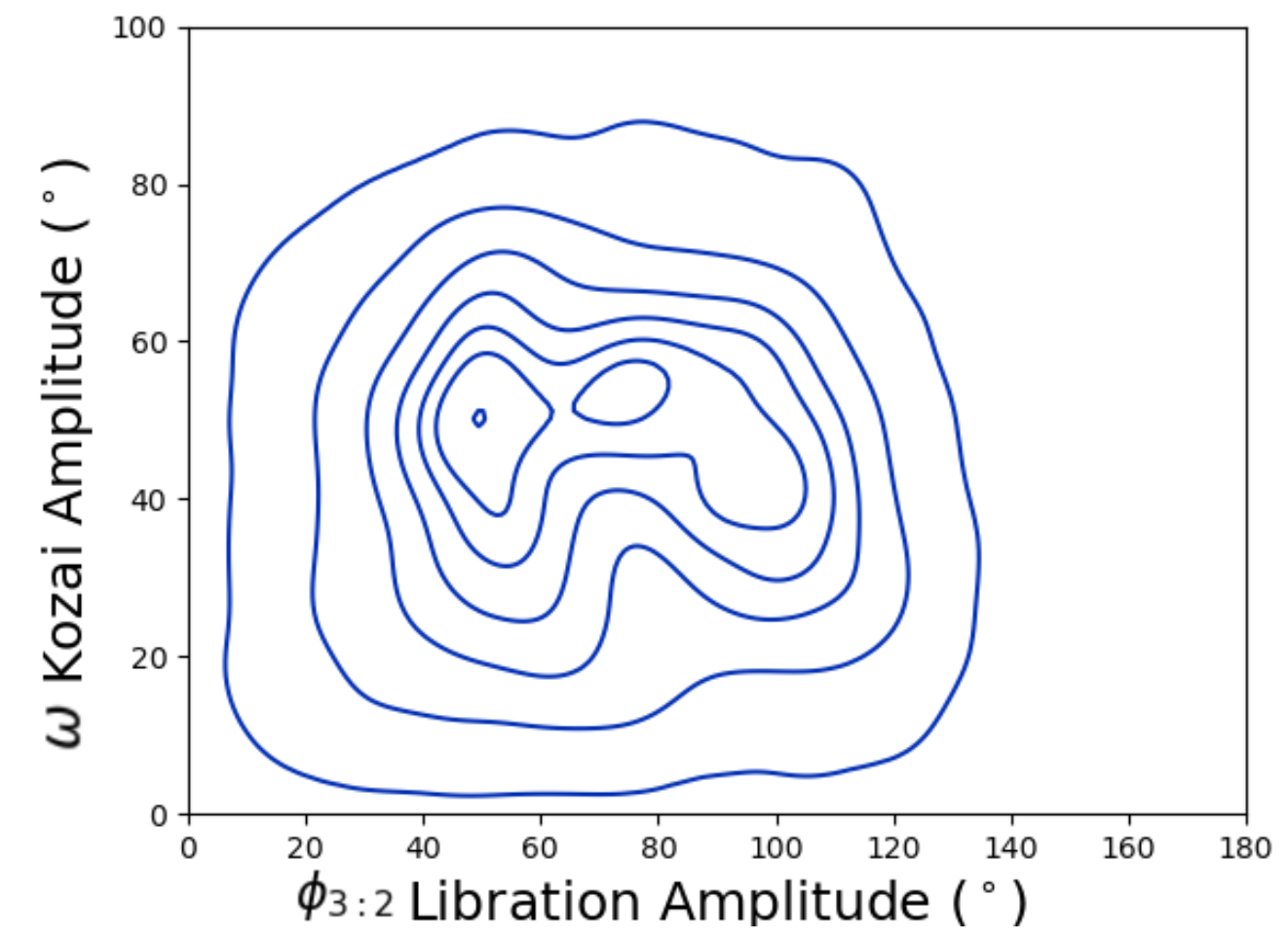}
    \caption{Nested contour lines or shades enclose 0.1\%, 10\%, 30\%, 50\%, 70\%, 90\%, and 99\% of 4-Gyr stable synthetic distribution Plutinos  in $a$ (top left) and $e$ (top right) versus $\phi_{3:2}$-libration amplitudes of the particles. 
    The distribution for the Kozai Plutinos is shown with blue contours, while the non-Kozai Plutinos are shown in reddish solid contours. 
    The lower panel shows the $\omega$-libration amplitudes versus the $\phi_{3:2}$-libration amplitudes for 4~Gyr stable Kozai Plutinos.
    }  
    \label{fig:libamps}
\end{figure}

Figure~\ref{fig:libamps} shows the density contours of the distribution of 4~Gyr-stable Plutino test particles, in osculating semimajor axis, eccentricity, and $\phi_{3:2}$-libration amplitude.
As shown in previous works \citep[e.g.,][]{Nesvorny2000,Tiscareno2009}, there is a clear pattern in that semi-major axes farther from the resonance center are required to have a larger $\phi_{3:2}$-libration amplitude. 
There is not a significant difference between the $\phi_{3:2}$-libration amplitude distribution for the Kozai and non-Kozai Plutinos - although the highest libration amplitudes ($\gtrsim$130$^{\circ}$) are only present in the non-Kozai Plutinos, possibly because the Kozai Plutino parameter space only includes the highest eccentricities ($e\gtrsim0.25$), and the largest $\phi_{3:2}$-libration amplitudes in the non-Kozai Plutinos tends to occur at lower $e$. 

There are some interesting density structure differences between the Kozai and non-Kozai Plutinos in this parameter space.  
The highest densities (thus, highest probability in parameter-space) of non-Kozai Plutinos occur at semimajor axes that are on either side of the resonance center, at $\phi_{3:2}$-libration amplitudes peaking around 100$^{\circ}$, and eccentricities centered around $e\sim0.2$.
Meanwhile, the Kozai Plutinos peak in density right on the resonance center at $a\sim39.4$~AU, and at higher eccentricities $e~\sim0.3$.
But the $e$ versus $\phi_{3:2}$-libration amplitude distribution  shows two peaks in density for the Kozai Plutinos, one around $\phi_{3:2}\sim100^{\circ}$ like the non-Kozai Plutinos, but another slightly lower density peak at $\phi_{3:2}\sim60^{\circ}$. 
The objects that have $\phi_{3:2}\sim60^{\circ}$ have a smaller $a$-range and larger $e$-range, while the objects with $\phi_{3:2}\sim100^{\circ}$ have a higher $a$-range and smaller $e$-range, causing the peaks to be at different amplitudes in the top two plots of Figure~\ref{fig:libamps}.

\subsection{Kozai Island Comparison and Kozai Parameter Space}
\label{sec:kozai}

Understanding the extent of the Kozai island parameter space as well as the characteristics of the trapped particles in the filled-parameter-space will provide useful insights for future modeling efforts and interpretation of real discoveries.
Characterizing the distribution of stable $\phi_{3:2}$-libration amplitudes for particles in Kozai as well as the amplitude of oscillation in $\omega$ is required.
It is also critically important to understand whether any of these parameters are interdependent in a filled parameter space synthetic distribution.
Any significant deviations between the real objects and a filled parameter space synthetic distribution would be indicative of emplacement mechanisms which preferentially populate the resonance with specific orbital characteristics.

For the Kozai Plutinos, we examine the distribution of $\phi_{3:2}$-libration and libration of $\omega$.
The bottom right panel of Figure~\ref{fig:libamps} shows that there is no significant correlation between the $\phi_{3:2}$-libration amplitude and $\omega$-libration amplitude for Kozai Plutinos; the full range of $\omega$-libration amplitudes correspond to the full range of $\phi_{3:2}$-libration amplitudes.  
These two libration modes appear to be independent from each other.
%Figure~\ref{fig:kozai_aei} shows that, 
As expected, we obtain no significant difference between the $a-e-i$ stable parameter space of the K90 (Kozai Plutinos that librate around $\omega=90^{\circ}$) and K270 (Kozai Plutinos that librate around $\omega=270^{\circ}$) islands.  
In our filled-parameter-space simulations, approximately the same number of long-term stable Kozai Plutinos remain in each island for the full 4~Gyr (3740 in K90 and 3788 in K270), with indistinguishable distributions of orbital elements.
This is not at all an unexpected result, and provides a nice check on our parameter-space-filling and integration technique.

Our integrations show that the vast majority (96\%) of Kozai Plutinos that experience $\omega$-libration for 4~Gyr remain librating in the same Kozai island for the duration of the simulation (Figure~\ref{fig:plutino_integrations} shows an example of $\omega$ and $\phi_{3:2}$ for a particle with this behaviour).   
Our simulations' prediction of most long-term stable Kozai Plutinos remaining in the same island into which they were originally captured provides some interesting diagnostic potential, and if the full \LIDO\ Survey results find that once biases are accounted for, the observed Kozai Plutinos are not consistent with a symmetric distribution, it could be very useful for testing migration models.

Our simulations revealed some unusual test particle behaviours.
We found a small subset of Kozai Plutinos that flip between $\omega$ libration islands while remaining stable Kozai Plutinos (which we will refer to as ``Kozai flippers''). 
Only about 4\% of 4~Gyr stable Kozai Plutinos show this behaviour.
Figure~\ref{fig:plutino_integrations} shows an example test particle that flips from being a K270 librator to a K90 librator of similar $\omega$-libration amplitude.
This particle transitions to a slightly larger $\phi_{3:2}$-libration amplitude before the $\omega$-libration flip occurs.
Another observed Plutino behaviour is unstable Kozai libration, which we do not classify as part of the Kozai Plutinos, although these test particles' orbits are clearly affected by $\omega$-libration.
Figure~\ref{fig:plutino_integrations} shows an example of an unstable Kozai librator, with large $\omega$-libration that alternates between librating around the K90 and K270 islands over the course of the whole simulation, with some periods of $\omega$-libration that are too large to remain in one island (although the higher density of points close to $0^{\circ}$ and 180$^{\circ}$ during these periods shows that there is something similar to $\omega$-libration still occuring).
Both Kozai flippers and unstable Kozai librators are sufficiently rare orbital configurations among the Plutinos (0.4\% and 6\% of all 4 Gyr stable Plutinos, respectively) that we do not expect to find real TNOs exhibiting this behaviour; long-term stable TNOs are not expected to be likely to move between Kozai islands.

The stable Plutinos exhibit both long-term and short-term stability of Kozai oscillation.
The unstable Kozai librators and Kozai flippers clearly inhabit very specific parts of Plutino $a-e-i$ parameter space in Figure~\ref{fig:kozai_flipaei}.
The unstable Kozai librators are all at the lowest $e$ and $i$ values possible for Kozai libration, while the Kozai flippers appears to straddle the line between the stable Kozai librators and unstable Kozai librators.

\begin{figure}%[!hp]
    \centering \includegraphics[width=0.5\textwidth]{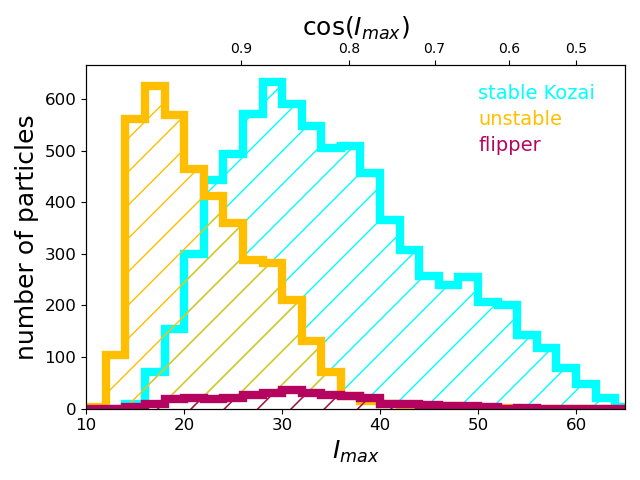}
    \caption{Histogram of $I_{\rm max}$ (see Eq.~\ref{eq:imax}) for all Kozai Plutinos in the simulation ($\cos I_{\rm max}$ along top axis for reference).  Kozai particles with different levels of stability have different $I_{\rm max}$ values: the lowest $I_{\rm max}$ values tend to be unstable Kozai librators (yellow), while the highest $I_{\rm max}$ values tend to be stable Kozai librators (cyan), with Kozai flippers inhabiting mid-range $I_{\rm max}$ values (magenta).
    }  
    \label{fig:imaxhist}
\end{figure}

In a simplified and properly averaged system,  the $z$-component of angular momentum $L_z\propto\cos~i\sqrt{1-e^2}$ is conserved.  Even in more general system, this can be approximately conserved  over the course a Kozai oscillation.   As in previous works \citep[e.g.,][]{Wan2007,Lawler2013}, different values of $L_z$ can be approximately parameterized for a given $e$ and $i$ of a Kozai particle at any timestep using $I_{\rm max}$, defined as
\begin{equation}
\cos~i\sqrt{1-e^2}=\cos~I_{\rm max}
\label{eq:imax}
\end{equation}
$I_{\rm max}$ is the inclination a particle would have when it reaches $e=0$, a useful shorthand for the ``energy in angular momentum."  
Of course, this will never happen for Kozai-cycling particles, because they need to have at least moderate $e$ and $i$ values - this is merely a convenient parameterization.
Figure~\ref{fig:imaxhist} shows the values of $I_{\rm max}$ for all Kozai Plutinos in our simulation set, which shows similar trends for the Kozai flippers and unstable Kozai librators as was seen in Figure~\ref{fig:kozai_flipaei}.
Each value of $I_{\rm max}$ can be though of as a different ``slice'' through $e\cos\omega-e\sin\omega$ parameter space, with a different range of $e$, $i$, and $\omega$ values that are possible over the course of Kozai cycles (see Figure~2 in \citet{Wan2007} and Figure~6 in \citet{Lawler2013} for ordered examples of these different parameter space shapes.)

Another way to look at these behaviours is in a different parameter space that better shows how Kozai cycles preserve angular momentum \citep[see, e.g.][]{Wan2007}.
For each of these particles, $e$ and $\omega$ change together in such a way over time that the particle stays librating around an $\omega$ island (either $\omega=90^{\circ}$ or $270^{\circ}$), although with slightly different $\omega$-libration amplitudes and $e$ ranges over the course of 4~Gyr due to small perturbations from Jupiter, Saturn, and Uranus.
Parameterization such as these are extremely useful for modelling resonant behavior in order to diagnose observation biases and population measurements. 
More recent works such as \citet{Lei2022} use an adiabatic approximation that is more accurate than Hamiltonian level surfaces sometimes used in the literature. 

\begin{figure}%[!hp]
    \centering \includegraphics[width=0.5\textwidth]{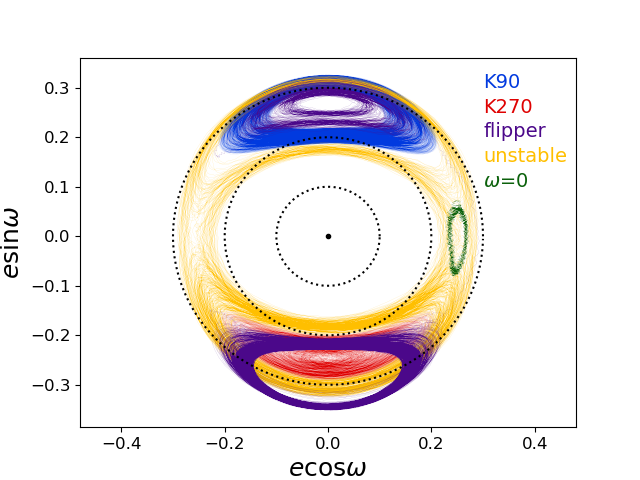}
    \caption{
    Selections from the 4~Gyr integration of several test particles are plotted in $e\cos\omega-e\sin\omega$, with example test particles exhibiting K270 libration (red), K90 libration (blue), unstable Kozai libration (yellow), Kozai flipping (magenta), and an example K0 librator in green, discussed in Section~\ref{sec:zero180}. 
    }  
    \label{fig:levelsurf}
\end{figure}

Several different example Kozai behaviours in are shown in Figure~\ref{fig:levelsurf}, a polar plot of $e$ and $\omega$ for selections from the 4~Gyr stability simulations for a few different test particles.  The stable K90 and K270 librators follow fairly low amplitude bean-shaped paths around $\omega=90^{\circ}$ and 270$^{\circ}$, as does the example flipper. 
The unstable Kozai Plutino has a large enough $\omega$ libration amplitude that it semi-regularly travels between $\omega$ islands.
The unstable Kozai Plutino behaviour appears to be exhibited by particles that are very close to the separatrix between Kozai and non-Kozai, and happen preferentially at the lower values of $I_{\rm max}$ (yellow histogram in Figure~\ref{fig:imaxhist}) as well as the lowest $e$ and $i$ portions of the stable phase space (yellow points in Figure~\ref{fig:kozai_flipaei}.  The Kozai flippers span a larger range of both $I_{\rm max}$ and $e-i$ and are likely more of a chance perturbation event - further investigation will be required with the real Kozai Plutinos discovered by \LIDO.

%%%%%%%%%%%%%%%%%%%%%%%%%%%%%%%%%%%%%%%%
\subsection{Outliers: 0$^{\circ}$ and 180$^{\circ}$ $\omega$-librators}
\label{sec:zero180}

Almost all of the Kozai Plutinos in our filled parameter space synthetic distribution librate around $\omega=90^{\circ}$ or $270^{\circ}$, but a handful librate around $\omega=0^{\circ}$ or $180^{\circ}$ for a large fraction of the simulation.  
No test particle librates in these modes for the full 4~Gyr simulations, but four Kozai Plutino particles (0.05\% of Kozai Plutinos in our simulation) show libration around $\omega=0^{\circ}$ or $180^{\circ}$ for up to 3~Gyr; one example K0 librator test particle is plotted in green in Figure~\ref{fig:levelsurf}.
Kozai libration around $\omega=0^{\circ}$ or $180^{\circ}$ has been previously reported in the literature for Plutinos \citep[and other resonances too, e.g.,][]{Lykawka2007}.
What may actually be happening here is libration around a small stable island that opens up in the adiabatic invariant curves for a particular set of orbital parameters \citep[e.g.][]{Lei2022}.  A future paper will explore this more accurate representation of the dynamics in detail for the Kozai Plutinos discovered by the \LIDO\ Survey.
The extremely small number of $0^{\circ}$ and $180^{\circ}$ Kozai librators we observe in our filled-parameter-space simulations leads us to predict that among real TNOs with high-quality orbits, there will be no $0^{\circ}$ and $180^{\circ}$ Kozai Plutinos until many thousands of Plutinos are known.

%%%%%%%%%%%%%%%%%%%%%%%%%%%%%%%%%%%%%%%%
\section{Discussion: Comparison with Neptune Migration Models}
\label{sec:Discussion}

The fraction of Plutinos that are in Kozai may have some diagnostic power for the mode of Neptune's migration, as well as the dynamical state of the proto-Kuiper Belt.
\citet{Lawler2013} compared theoretical predictions and observational measurements of the Kozai fraction \fkoz\ that had been measured up until that point in time.
Theoretical predictions from capture simulations include 20-30\% from smooth migration \citep{Chiang2002}, 19\% from smooth migration \citep{Hahn2005}, and 16\% from a Nice model simulation \citep{Levison2008}.
Observational (biased) measurements of \fkoz\ include 8\% \citep{Gladman2012},  $\sim$30\% \citep{Lykawka2007}, 33\% \citep{Schwamb2010}, and 24\% \citep{Volk2016}.
Debiased measurements of \fkoz, requiring well-characterized surveys, have so far been published only by OSSOS++ surveys, giving 95\% confidence ranges of \fkoz$<33\%$ \citep{Gladman2012} and $0.08<$\fkoz$<0.35$ \citep{Volk2016}.

The $\LIDO$ analysis of the discovered and debiased Plutinos and Kozai Plutinos is in preparation and will be presented in future works.  For a preliminary, observation-bias-free comparison to our filled-phase space sample, here we conduct a reanalysis of the Neptune migration simulations from \citet{Kaib2016}, which were dynamically classified in \citet{Lawler2019}.
Previous work has replicated survey biases and compared these simulated TNOs to e.g., real TNOs just sunward of resonances \citep{Pike2017,Lawler2019,Bernardinelli2022}, and found that grainy slow Neptune migration provides a compelling match to the resonant dropout supbpopulations of the outer solar system.
Major advantages to using the results of planetary migration simulations include the larger sample size, orbit fits without uncertainty, and no observation-bias imposed due to apparent magnitudes of the synthetic objects.
These simulations were set up to compare the effects of migration speed and graininess on the dynamical structure of the Kuiper Belt, based on a ``jumping Jupiter''-style migration \citep[e.g.,][]{Brasser2009}.  
These simulations use two different Neptune migration speeds, a faster 30~Myr post-jump $e$-folding timescale, and a slower 100~Myr timescale.  
They also use two different migration modes: standard smooth migration, and grainy migration, where Neptune's outward migration includes small random jumps in $a$ that simulate the scattering of $\sim$Pluto-sized planetesimals \citep[e.g.,][]{Nesvorny2016}. These two migration speeds and modes yield four separate simulations, which we will refer to as GF (grainy fast), GS (grainy slow), SmF (smooth fast), and SmS (smooth slow).
Continuing from the dynamical classification of these models presented in \citet{Lawler2019}, we searched the 10~Myr dynamical simulations specifically for Kozai Plutinos and analysed their properties, looking for any differences between the simulations.

\begin{figure}%[!hp]
    \includegraphics[width=0.5\textwidth]{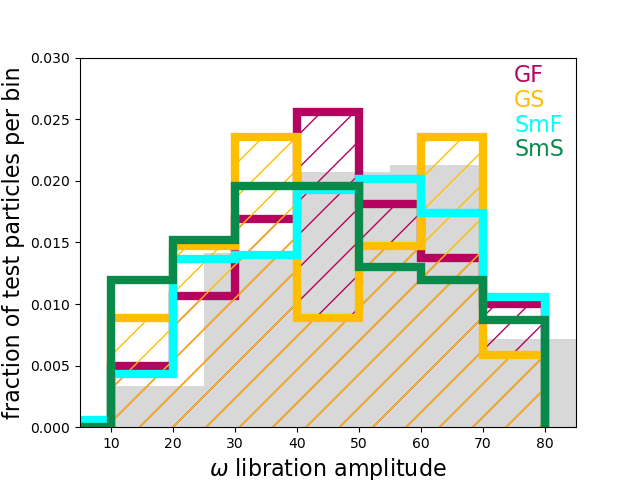}
    \includegraphics[width=0.5\textwidth]{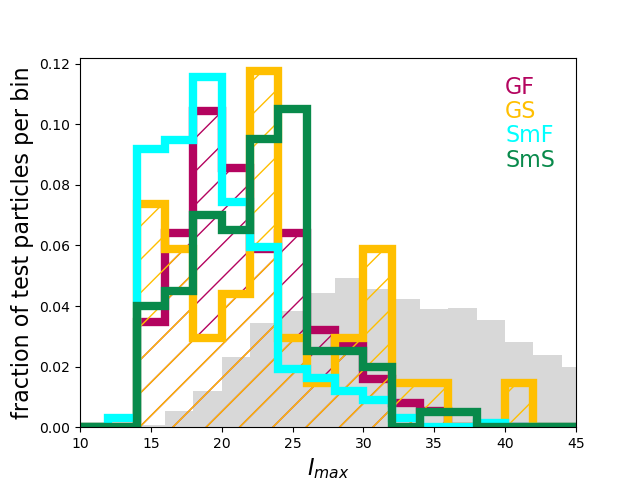}
    \caption{The distribution of $\omega$-libration amplitudes and $I_{\rm max}$ values resulting from the test particles in the four simulations analysed from \citet{Kaib2016} (see legend), compared with the same distributions for our filled-parameter-space synthetic distribution (grey filled histogram). Despite different Neptune migration timescales and smooth or grainy migration, these four simulations result in similar distribution ranges of $\omega$-libration while being very different in $I_{\rm max}$.
    }  
    \label{fig:KSkoz}
\end{figure}

We were specifically interested in the characteristics of the Kozai Plutinos captured in the different migration simulations.
Figure~\ref{fig:KSkoz} shows the $\omega$-libration amplitude and $I_{\rm max}$ distributions for each of the four simulations, and Table~\ref{tab:KSkoz} summarizes these distributions.
In order to compare the 4 simulations, we calculate the bootstrapped Anderson-Darling statistic \citep{AndersonDarling} for each simulation compared with the other simulations, in the distributions of $\omega$-libration amplitudes and $I_{\rm max}$ values.
In most of the $\omega$-libration amplitude comparisons, the bootstrapped AD values do not allow rejection of the null hypothesis that one distribution could be drawn from the other (this means the two distributions are statistically the same).
But the GF $\omega$-libration amplitude is inconsistent with the other three $\omega$-libration amplitude distributions. 
The $I_{\rm max}$ distributions seem to be more tightly distributed and thus more inconsistent with each other - only the GS $I_{\rm max}$ distribution is consistent with any of the other $I_{\rm max}$ distributions, and this may be because of the small number of test particles (35).
So, as expected, there are significant differences between the properties of the Kozai Plutinos captured by each of the 4 simulations, though it is not clear from the analysis here if graininess or migration timescale has a stronger effect on different aspects of the Kozai Plutino parameter space distribution.

These emplacement simulation distributions may be compared with the same distributions from our filled-parameter-space synthetic distribution, shown in grey.
We observe that emplacement by Neptune migration results in much lower $I_{\rm max}$ values than the filled-parameter-space synthetic distribution, but similar distributions of $\omega$-libration amplitudes; this difference could be due to the fact that the filled-parameter-space synthetic distribution includes higher inclinations than are likely to exist in the real Solar System.
We note that our filled-parameter-space synthetic distribution histograms include only 4~Gyr stable Kozai librators, while the \citet{Kaib2016} simulations are diagnosed with shorter 10~Myr integrations (after the initial 4~Gyr migration simulations), so some long-term unstable Kozai Plutinos are no doubt included in those distributions, but that alone is not enough to explain the difference in $I_{\rm max}$ distributions that we observe - only 6\% of the filled-parameter-space Plutinos are unstable Kozai.

\begin{deluxetable}{c|c|ccc|cc|cc}
\tablecaption{$\omega$ libration amplitudes and $I_{\rm max}$ for this work and Neptune migration models \label{tab:KSkoz}}
\tablewidth{0pt}
\tablehead{Simulation & $f_{\rm koz}$ & \multicolumn{3}{c}{$N_{\rm koz}$} & \multicolumn{2}{c}{avg. $\omega$ lib.\ amp.\ } & \multicolumn{2}{c}{avg.\ $I_{\rm max}$} \\
 & & K90 & K270 & other & K90 & K270 & K90 & K270
}
\startdata
GF	&	34\%	&	96	&	87	&	5	&	52$^{\circ}$	$\pm$	21$^{\circ}$	&	55$^{\circ}$	$\pm$	20$^{\circ}$	&	23$^{\circ}$	$\pm$	5$^{\circ}$	&	21$^{\circ}$	$\pm$	4$^{\circ}$	\\
GS	&	42\%	&	15	&	19	&	0	&	40$^{\circ}$	$\pm$	17$^{\circ}$	&	48$^{\circ}$	$\pm$	17$^{\circ}$	&	24$^{\circ}$	$\pm$	7$^{\circ}$	&	23$^{\circ}$	$\pm$	6$^{\circ}$	\\
SmF	&	43\%	&	168	&	167	&	2	&	51$^{\circ}$	$\pm$	20$^{\circ}$	&	49$^{\circ}$	$\pm$	17$^{\circ}$	&	20$^{\circ}$	$\pm$	4$^{\circ}$	&	20$^{\circ}$	$\pm$	4$^{\circ}$	\\
SmS	&	59\%	&	48	&	52	&	0	&	44$^{\circ}$	$\pm$	21$^{\circ}$	&	48$^{\circ}$	$\pm$	20$^{\circ}$	&	23$^{\circ}$	$\pm$	5$^{\circ}$	&	22$^{\circ}$	$\pm$	4$^{\circ}$	\\ \hline
this work: all	&	10.8\%	&	3740	&	3788	&	$-$	& 	35$^{\circ}$	$\pm$	10$^{\circ}$	&	35$^{\circ}$	$\pm$	10$^{\circ}$	&	49$^{\circ}$	$\pm$	17$^{\circ}$	&	49$^{\circ}$	$\pm$	16$^{\circ}$	\\
this work: $i<50^{\circ}$	&	16.9\%	&	3479	&	3532	&	$-$	& 	34$^{\circ}$	$\pm$	9$^{\circ}$	&	34$^{\circ}$	$\pm$	9$^{\circ}$	&	50$^{\circ}$	$\pm$	17$^{\circ}$	&	50$^{\circ}$	$\pm$	16$^{\circ}$	\\
this work: $i<40^{\circ}$	&	15.5\%	&	2884	&	2835	&	$-$	& 	31$^{\circ}$	$\pm$	7$^{\circ}$	&	31$^{\circ}$	$\pm$	6$^{\circ}$	&	52$^{\circ}$	$\pm$	16$^{\circ}$	&	51$^{\circ}$	$\pm$	15$^{\circ}$	\\
\enddata
\tablecomments{Note that values from this work are for 4~Gyr stable Kozai Plutinos only, while the four simulations from \citet{Kaib2016} are only run for 10~Myr, so likely include some Kozai Plutinos that are not long-term stable. Error bars are standard deviation of the mean.}
\end{deluxetable}	

The four Neptune migration simulations that are analysed here show no significant differences between any of the captured orbital distributions between the K90 and K270 islands (Table~\ref{tab:KSkoz}), and all four start with the same pre-sweeping test particle distribution.
Simulations in \citet{Li2014} include Neptune's 3:2 resonance sweeping through populations with different levels of excitation, finding that TNOs with high pre-capture inclinations (30-40$^{\circ}$) are preferentially trapped in the K90 island.
If the initial population of TNOs was dynamically excited prior to being swept into the Plutino resonance, we expect to find a smaller population in the K270 $\omega$-libration island than around K90 island, and overall we expect to see a broader inclination distribution in the K90 $\omega$-librators.

We do not expect \fkoz\ for the filled-parameter-space synthetic distribution presented here (\fkoz=10.8\%) to match reality, since we have no reason to think that Plutino parameter space was completely filled during Neptune migration. 
But measured deviations from the \fkoz\ filled-parameter-space synthetic distribution should tell us about the Plutino capture process, and these four Neptune migration simulations are a first step to testing this.
While the differences between the $I_{\rm max}$ and $\omega$-libration distributions are small between the four migration models (though significantly different between some models), the Kozai fraction \fkoz\ has a large amount of variation with different speeds and migration modes.
The highest \fkoz\ results from slow smooth migration, with nearly 60\% of Plutinos in Kozai at the end of this migration simulation, while the smallest fraction (34\%) of Plutinos were in Kozai in the grainy fast migration.
This suite of simulations shows that grainy migration appears to be approximately 75\% as efficient at capturing Kozai Plutinos, and faster migration appears to be about 75\% as efficient again.

Measuring \fkoz\ to high precision will be a powerful tool for constraining Neptune's migration.
However, it is important to keep in mind that the extensive observational bias modelling in \citet{Lawler2013} shows that \fkoz\ cannot be measured in survey data unless the survey is well-characterized, and survey pointing locations, magnitude limits, and tracking fractions are measured.  
One of the significant challenges in previous works was constraining the inclination distribution of the Kozai Plutino population, as this is poorly sampled due to the on-ecliptic nature of the surveys and the small sample size.
Following the framework of the careful OSSOS++ survey biasing-adjusted modelling for the Plutinos \citep{Kavelaars2009, Gladman2012, Alexandersen2016, Volk2016}, we expect similar well-constrained results from the off-ecliptic \LIDO\ Survey in future papers.

\section{Conclusions}
\label{sec:Conc}

In this paper we present a filled-parameter-space synthetic distribution of the Plutinos and discuss the properties of 4~Gyr stable Plutinos and Kozai Plutinos. 
\citet{Gomes2000} predicted that the Kozai resonance would be important to the distribution of Plutinos today:   
the intrinsic fraction of the Plutinos in Kozai resonance is a result of the specifics of the planetary migration which populated the outer Solar System.  
A detailed synthetic distribution of the available parameter space provides a useful tool for interpreting both the results of planetary migration simulations and TNO discoveries, and our synthetic distribution has been made available at \url{https://www.canfar.net/citation/landing?doi=23.0028}.

One critical step in assessing the diagnostic power of comparing Kozai to non-Kozai Plutinos is to determine whether any signature imparted during planetary migration would last until the present day.
Our integrations show that $\sim$96\% of 4~Gyr stable Kozai Plutinos stay in the same island for the duration of the simulation. 
The fact that long-term stable Kozai Plutinos stay in their original $\omega$-libration island may provide an interesting and perhaps powerful diagnostic: if, as suggested by simulations in \citet{Li2014}, the initial fraction of Plutinos captured into each Kozai island depends on the the excitation of the population that Neptune's 3:2 resonance sweeps across during Neptune's outward migration, then any asymmetry in the number of TNOs that today inhabit the two Kozai islands provides constraints on the initial dynamical state of the Kuiper Belt prior to Neptune's migration.
Capture via scatter is expected to dominate the emplacement of plutinos \citep[e.g.][]{Pike2023}, erasing any information about the initial eccentricity distribution of the disk.  
However, the population fraction and inclination distribution of Kozai Plutinos in the K90 and K270 islands is a signature of the initial inclination distribution of the proto-planetesimal disk, testable by real plutino discoveries.

Re-analysis of the planetary migration simulations in  \citet{Lawler2019} shows that different Neptune migration modes (fast/slow, grainy/smooth) do not significantly affect the K90/K270 fractions, which remain at $\sim$50\% in each Kozai libration island. 
However, the total fraction of Kozai to non-Kozai Plutinos is significantly higher for smooth migration than for grainy, and is higher for slower Neptune migration speeds.
This suggests that K90/K270 population comparisons constrain the initial disk distribution, and the total Kozai/non-Kozai fraction constrains the mode of Neptune's migration.

In order to measure \fkoz\ and any asymmetry in Kozai Plutino libration islands, a number of Kozai Plutinos need to be discovered in well-characterized surveys, which can be used to account for extensive, complex observational biases in the discovery of Kozai Plutinos \citep{Lawler2013}.
With 20 medium-high inclination preliminary Plutinos discovered by \LIDO\ (see Figure~\ref{fig:realplut_aei}), we expect that \LIDO\ will place more strict constraints on \fkoz\ and Kozai Plutino asymmetries, particularly at smaller sizes than future wide-field surveys will be sensitive to.
The Vera C. Rubin Observatory's Legacy Survey of Space and Time is expected to detect thousands of new TNOs over the next decade, with known observational biases and a powerful Survey Simulator \citep{Schwamb2023} that will help constrain the portion of stable Plutino space that was populated during the giant planet migration era. 

\vspace{1cm}

The authors acknowledge the sacred nature of Maunakea and appreciate the opportunity to observe from the mountain. CFHT is operated by the National Research
Council (NRC) of Canada, the Institute National des Sciences de l’Universe of the Centre National de la Recherche Scientiﬁque (CNRS) of France, and the University of Hawaii, with \LIDO\ receiving additional access due to contributions from the Institute of Astronomy and Astrophysics, Academia Sinica, Taiwan. Data were produced and hosted at the Canadian Astronomy Data Centre; processing and analysis were performed using computing and storage capacity provided by the Canadian Advanced Network For Astronomy Research
(CANFAR), operated in partnership by the Canadian Astronomy Data Centre and The Digital Research Alliance of Canada with support from the National Research Council of Canada the Canadian Space Agency, CANARIE and the Canadian Foundation for Innovation.
This paper includes data gathered with the 6.5 meter Magellan Telescopes located at Las Campanas Observatory, Chile.

The authors wish to acknowledge the land on which they live and carry out their research: Canadian Treaty 4 land, which is the territories of the n\^{e}hiyawak, Anih\v{s}in\={a}p\={e}k, Dakota, Lakota, and Nakoda, and the homeland of the M\'{e}tis/Michif Nation. Center for Astrophysics | Harvard \& Smithsonian is located on the traditional and ancestral land of the Massachusett, the original inhabitants of what is now known as Boston and Cambridge. We pay respect to the people of the Massachusett Tribe, past and present, and honor the land itself which remains sacred to the Massachusett People.

This research has been supported in part by NSERC Discovery Grant RGPIN-2020-04111 (SML). REP, MA, and CC acknowledge NASA Solar System Observations grant 80NSSC21K0289. CC was supported in part by a Massachusetts Space Grant Consortium (MASGC) Award.

We thank Nate Kaib for providing the simulation output that was used by the \citet{Lawler2019} paper and by extension this paper. We also thank X.-S. Wan and T.-Y. Huang for providing the disturbing function coefficients for Kozai Plutinos.

\facilities{CFHT, CANFAR, Magellan} 
\software{This research was made possible by the open-source projects 
\texttt{REBOUND} \citep{rein2012},
\texttt{WHFast} \citep{Rein2015},
\texttt{Jupyter} \citep{Kluyver2016},
\texttt{matplotlib} \citep{Hunter2007,Droettboom2016},
\texttt{numpy} \citep{Harris2020},
the updated version of the CFEPS Moving Object detection Pipeline \citep{Petit2004} used by OSSOS,
\texttt{MegaPipe} \citep{Gwyn2008}}

\bibliographystyle{aasjournal}
%\bibliography{export-bibtex}

\end{document}